\documentclass[runningheads, anonymous]{llncs}
\usepackage{float}

\usepackage{cite}
\usepackage{amsmath,amssymb,amsfonts}
\usepackage{algorithmic}
\usepackage{graphicx}
\usepackage{textcomp}
\usepackage{listings}
\usepackage{multicol}
\usepackage{makecell}
\usepackage{longtable}
\usepackage{xcolor}
\usepackage{paralist}
\usepackage{multirow}
\usepackage{booktabs}
\usepackage{url}
\usepackage{soul}
\usepackage{xspace}
\usepackage{wrapfig}
\usepackage{hyperref}
\newcommand{\tool}[1]{\textsc{#1}\xspace}
\newcommand{\toolname}{\tool{SyzAgent}}

\newcommand{\commentout}[1]{}

\usepackage{tcolorbox}
\usepackage{marginnote}

\usepackage[T1]{fontenc}
\usepackage{graphicx}
\usepackage{comment}
\begin{document}
\title{Towards Large Language Model Guided \\Kernel Direct Fuzzing }

\authorrunning{F. Author et al.}
\institute{Anonymous Institutes
}
\author{Xie Li\inst{1,2}
\and
Zhaoyue Yuan\inst{1,2}\and
Zhenduo Zhang\inst{4}\and\\
Youcheng Sun\inst{3,5}\and
Lijun Zhang\inst{1,2,4}}
\institute{
Key Laboratory of System Software (Chinese Academy of Sciences) and 
 State Key Laboratory of Computer Science,
Institute of Software, Chinese Academy of Sciences 
\and 
University of Chinese Academy of Sciences
\and
MBZUAI
\and
Automotive Software Innovation Center, Chongqing, China
\and
The University of Manchester
}
\maketitle              
\begin{abstract}
Direct kernel fuzzing is a targeted approach that focuses on specific areas of the kernel, effectively addressing the challenges of frequent updates and the inherent complexity of operating systems, which are critical infrastructure. This paper introduces \toolname,
a framework that integrates LLMs with the state-of-the-art kernel fuzzer Syzkaller, where the LLMs are used to guide the mutation and generation of test cases in real-time.
We present preliminary results demonstrating that this method is effective on around 67\% cases in our benchmark during the experiment.

\end{abstract}
\section{Introduction}

Operating systems (OS) are crucial in modern computing infrastructures, making the correctness and reliability of an OS kernel vital. Fuzzing is a common method for identifying software vulnerabilities and has been notably applied in kernel testing with tools like Syzkaller \cite{syzkaller}, which has identified many bugs. Despite progress, the complexity of modern OSes can impede fuzzers from reaching deeper code paths. To improve fuzzing efficiency and coverage, researchers have explored ways to better discover and utilize the dependency relations between system calls and tasks \cite{healer, rtkaller}. Other works have employed reinforcement learning techniques \cite{syzvegas} and static analysis methods \cite{difuze, statefuzz} to target previously unreached code during fuzzing. 

With the rapid advancement of generative AI \cite{achiam2023gpt}, the use of large language models (LLMs) in system fuzzing is increasingly recognized \cite{fuzz4all}. The KernelGPT method \cite{kernelgpt} has been proposed to utilizing LLMs to generate Syzlang, a domain-specific language for system calls, facilitating improved seed generation and test case creation in Syzkaller \cite{syzlang}.

Instead of general kernel fuzzing like Syzkaller, this work emphasizes direct kernel fuzzing, which targets specific, often critical areas within the OS kernel to manage the challenges posed by frequent updates and rapid iterations. The Syzdirect approach \cite{syzdirect} extends Syzkaller by leveraging the call graph and resource model to provide structured guidance for generating test cases more effectively, enabling the more effective direct kernel fuzzing.

In this work, we integrate LLMs with direct fuzzing of the OS kernel. The source code and fuzzing intermediate results are fed to the LLM dynamically to retrieve guidance for test case generation. Unlike KernelGPT, which focuses on generating Syzlang specifications, and Syzdirect, which utilizes pre-built guidance from the call graph and resource model, our approach employs real-time feedback from the LLM to adapt to changes in the kernel. We implemented our framework, \toolname, to achieve this integration and provide preliminary experimental results demonstrating the effectiveness of the approach. 
Without loss of generality, GPT-4o \cite{openai2023gpt4o} is used for the experiments in this paper. In addition, we share insights into the challenges and experiences encountered while integrating LLMs with kernel fuzzers.

\section{Motivating Example}
\label{sec:motivating}

Consider a commit changing the function \texttt{\_\_anon\_inode\_getfd} in the Linux kernel, referred to as the \emph{target function}. Our objective is to test the newly introduced code in this commit using guidance from a LLM.  

By compiling and analyzing the Linux kernel, we generate a set of \emph{call paths}, which represent the routes from system calls to the target function in the kernel's call graph. 
Below are two example call paths, where \texttt{func1} $\rightarrow$ \texttt{func2} indicates that \texttt{func2} is called within the body of \texttt{func1}:

\vspace{-16pt}
\begin{align*}
&\texttt{\footnotesize inotify\_init} \rightarrow \texttt{\footnotesize do\_inotify\_init} \rightarrow
\texttt{\footnotesize anon\_inode\_getfd} \rightarrow
\texttt{\footnotesize \_\_anon\_inode\_getfd}\\\vspace{10pt}
&\texttt{\footnotesize mock\_drm\_getfile} \rightarrow \texttt{\footnotesize anon\_inode\_getfile} \rightarrow
\texttt{\footnotesize \_\_anon\_inode\_getfd}\\
\end{align*}
\vspace{-30pt}

The first path illustrates a direct call to the target function from a system call (\texttt{inotify\_init}), while the second involves an indirect call through other functions within two steps.
We collect these paths to inform the LLM about potential triggers for the target function. Once identified, the source code from these paths, referred to as \emph{calling code}, will be extracted and used to formulate the initial prompt, as shown in Prompt~\ref{lst:init_prompt} \footnote{Detailed prompts is available at \url{https://github.com/SpencerL-Y/ChatAnalyzer/blob/main/chat_interface.py}.}.

\lstset{
    basicstyle=\small\ttfamily, 
    breaklines=true,
    breakindent=0pt,             
    showspaces=false,           
    xleftmargin=2em,             
    framexleftmargin=2em,           
    aboveskip=0.3em,             
    belowskip=0.3em,             
    lineskip=-2pt,             
    numberstyle=\tiny\color{gray}
}
{
\begin{lstlisting}[caption={Initial prompt (before fuzzing)},label={lst:init_prompt}]
[calling code]
Above is the source code that may call function [target function], which system calls may trigger the call path of function [target function]? 
\end{lstlisting}

Upon receiving the initial prompt, the LLM identifies the following system calls that may potentially interact with the target function.

\vspace{5pt}
\texttt{[inotify\_init, inotify\_init1, fsopen, fspick, perf\_event\_open,} \\
\texttt{timerfd\_create, epoll\_create, epoll\_create1, eventfd, eventfd2,} \\
\texttt{signalfd, signalfd4]}
\vspace{5pt}

These system calls represent possible entry points to the target function, and the reason to use the LLM for analysis is to leverage its potential to identify additional system calls, as LLM may provide more diverse outcome since it has been trained on extensive open-source project data. 

Subsequently, a kernel fuzzer like Syzkaller is launched with  generated test cases using  system calls with increased  probability to reaching the target. During the fuzzing process, whenever 500 test cases are executed, those that covering functions within 2 steps of the target function are collected, and the covered source code is recorded to create a feedback prompt, as shown in Prompt~\ref{lst:feedback_prompt}. 

\lstset{
    basicstyle=\small\ttfamily, 
    breaklines=true,
    breakindent=0pt,             
    showspaces=false,           
    xleftmargin=2em,             
    framexleftmargin=2em,        
    aboveskip=0.3em,             
    belowskip=0.3em,             
    lineskip=-2pt,                  
    numberstyle=\tiny\color{gray}
}

\begin{lstlisting}[caption={Feedback prompt (during fuzzing)},label={lst:feedback_prompt}]
[calling code]
Above is the source code that may call the target function [target function], in testing procedure we found the following system call programs reach functions that is close to the target function:
[test cases and covered code]
Generate a list of system calls that if the probability of generating such system calls in test cases is increased, the fuzzing process is more possible to reach our target function: [target function]?
\end{lstlisting}

After receiving the feedback prompt, the LLM provides an updated list of system calls. With real test cases available, the LLM is more likely to introduce related system calls. In this example, the LLM adds two more system calls, \texttt{drm\_syncobj\_handle\_to\_fd\_ioctl} and \texttt{mmap}, to the initial list. This improvement in system call generation allows the fuzzing process to cover the target function more frequently in subsequent runs.

\section{Approach}
The architecture of our proposed approach is depicted in Figure \ref{fig:framework}, comprising two parts: 1) the original kernel fuzzer Syzkaller, and 2) its LLM extension, \toolname. Below, we introduce each part and explain their interactions.

 \begin{figure}[h]
\vspace{-10pt}
     \centering
         \includegraphics[width=1\linewidth]{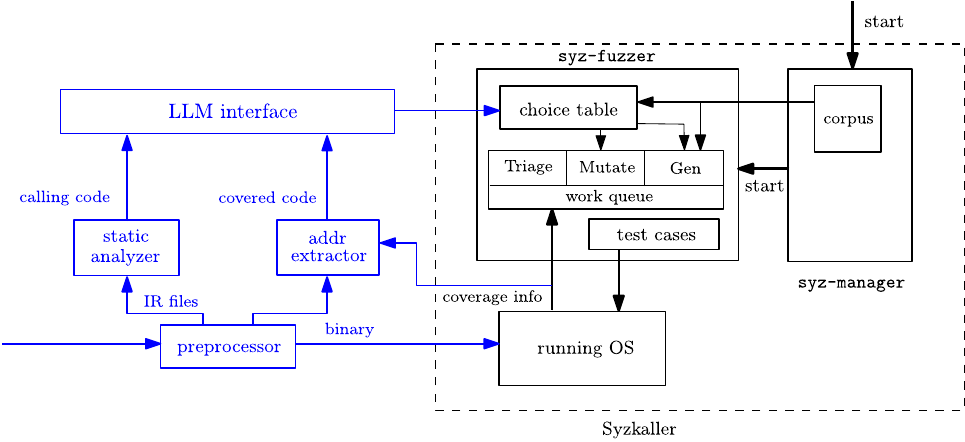}
     \caption{\toolname extends the existing Syzkaller by applying LLM in fuzzing kernels.}
     \label{fig:framework}
\vspace{-10pt}
 \end{figure}

\subsection{Syzkaller}

Syzkaller fuzzes the OS kernel by executing finite sequences of system calls with their arguments, where system call comes from a set of system calls $S$. It creates three task types in the work queue (as shown in Figure~\ref{fig:framework}):
\vspace{-5pt}
\paragraph{Generation} Initial seed programs are generated from manually tuned templates to ensure deeper test cases.
\vspace{-5pt}
\paragraph{Mutation} Mutation is applied to programs selected from a corpus (i.e., previously executed programs with new coverage). During this phase, system calls and their arguments are modified, including adding, removing, or changing system calls. This process is guided by the fuzzing state, which includes a choice table ($ct$): a two-dimensional array where $ct[c][c']$ represents the probability of generating system call \texttt{c}{$'$} after \texttt{c}. System call insertion is either random (5\% of the time) or based on probabilities from the choice table. Arguments are generated by considering available resources at the insertion point.
\vspace{-5pt}
\paragraph{Triage} Test cases that triggered new coverage will be verified and minimized by removing redundant system calls, with successful cases added to the corpus as new seeds. Triage tasks take priority in the fuzzing process, followed by generation and mutation if no triage tasks are available.

\subsection{\toolname}

We propose \toolname to extend Syzkaller, as shown in Figure~\ref{fig:framework}. It integrates an LLM into Syzkaller's generation and modification of the choice table. The LLM influences the fuzzing process in three key procedures: 
\begin{inparaenum} 
    \item It constructs the initial choice table based on static analysis and LLM analysis results.
    \item During fuzzing, it collects some running test cases with coverage information during fuzzing, formulates feedback prompt to obtain guidance on fuzzing from LLM.
    \item Finally it updates the choice table using guiding information provided by the LLM.
\end{inparaenum}
The extension corresponds to the four new components: 
the \emph{preprocessor}, \emph{static analyzer}, \emph{address extractor}, and \emph{LLM interface}, as shown in Figure~\ref{fig:framework}.


\vspace{-10pt}
\subsubsection{Pre-Processor} The pre-processor compiles the OS kernel's source code into a binary image for testing and generates intermediate representations (IR)  from LLVM framework\cite{llvm}.  These generated IR files are used for static analysis and are avoided from any optimization to reflect the calling relation as detailed as possible. Additionally, the pre-processor gathers information on all C functions present in the Linux kernel.
 
\vspace{-10pt}
\subsubsection{Static Analyzer} The static analyzer parses and analyzes the IR files generated by the pre-processor, resulting in the call graph of the OS kernel. A \emph{call graph} of the Linux kernel is a graph \( G = (C \cup F, E) \), where: \( C \) is a finite set of system calls, \( F \) is a finite set of other functions, and \( E \subseteq (C \cup F) \times (C \cup F) \) represents the set of directed edges in the call graph, showing the calling relationships between functions.
Given a target function $f_t$, the static analyzer performs following tasks: 
\begin{itemize}
    \item[\textbf{Job 1:}] Find all paths from some \( s \in C \) to \( f_t \). This corresponds to the first type of call paths in the  motivating example.
    \item[\textbf{Job 2:}] Find all paths from any function \( f \) to \( f_t \) with length \( l \), where \( l < \mathbf{k} \) and  \( \mathbf{k} \in \mathbb{N} \) is a constant. These are the second type of call paths in the \st{demo} motivating example;
    \item[\textbf{Job 3:}] Identify all \emph{close functions} \( f_c \) within a specific \emph{close range} constant \( \mathbf{d} \in \mathbb{N} \), where \( f_c \) is an \( n \)-step predecessor in the call graph and \( n \leq \mathbf{d} \). A predecessor refers to a function that directly calls or influences another function in a call path.
    These functions are denoted by \emph{close area}.
\end{itemize}

\vspace{-10pt}
\subsubsection{Address Extractor} The address extractor matches program counter (PC) points in the compiled Linux kernel binary to their actual locations. Syzkaller uses KCOV~\cite{kcov} for coverage feedback, tracking the PC points reached by test cases. To improve efficiency, PC points in the close area are extracted in advance for quicker coverage checks.

\vspace{-10pt}
\subsubsection{LLM-Interface} The LLM interface communicates with the LLM by sending the initial and feedback prompts. It then extracts a set of system calls, \( S_{inc} \), from the LLM feedback to update the choice table.

The choice table is modified as follows: for any system calls \( c_1 \) and \( c_2 \), let \( ct_0[c_1][c_2] \) represent the original choice table value in Syzkaller. The LLM-updated value, \( ct_1[c_1][c_2] \), is set to \( ct_0[c_1][c_2] + 1 \) if either \( c_1 \in S_{inc} \) or \( c_2 \in S_{inc} \); otherwise, \( ct_1[c_1][c_2] = ct_0[c_1][c_2] \). The final choice table is computed by normalizing \( ct_1 \) for each row:
$
ct[c_1][c_2] = \dfrac{ct_1[c_1][c_2]}{\sum_i ct_1[c_1][c_i]}.
$

Apart from components above, it is worth mentioning that since LLM analysis runs slower than fuzzing. We sample test cases from all test cases and run LLM analysis in parallel. 
For every 500 cases sampled, some cases that covered close area will be selected randomly to do the feedback prompting via LLM-interface.

\section{Preliminary Experimental Results}
We conducted experiments on fuzzing the Linux kernel to demonstrate that our LLM-driven \toolname method: 1) effectively adapts the existing vanilla Syzkaller tool, even breaking its coverage plateau, and 2) offers advantages over the specialized direct kernel fuzzing tool, SyzDirect.

Our experimental setup consisted of a PC equipped with a 13th Gen Intel Core i7-13700 processor and 128GB of memory. The virtual machine under test was configured on QEMU, running a Linux system on an AMD architecture with 4 CPUs and 4GB of memory. Given the 12.8k token limit of the LLM-interface in GPT-4o \cite{openai2023gpt4o}, we selected target functions based on the principle that no function in their call paths should have more than five predecessors to prevent the explosion of number of calling paths of the target function. From this set, we  selected a total of 27 target functions which our tool can process currently as our benchmark. 

\vspace{-10pt}
\paragraph{\toolname vs Syzkaller}
In this experiment, each target function was fuzzed using both \toolname and Syzkaller, with each tool tested three times per function, and each run limited to two hours. The fuzzing results from \toolname and Syzkaller are summarized in Table~\ref{tab:exp_table}. Out of the 27 cases, \toolname achieved a hit rate (the ratio of number of test cases hit close area and the number of all test  cases) that surpassed Syzkaller by more than $10\%$ in 8 cases, while it underperformed compared to Syzkaller in only 5 cases. Comparably, we also compute how the increased coverage outperforms the original one, represented as $\omega = \frac{\text{Avg. Diff}}{\text{Avg. Syzkaller Hit Rate}}$ and 18 cases out of 27 have $\omega \ge 10\%$ which is 67\%.

\begin{table}[]
\centering
\begin{tabular}{|c|p{3.5cm}|c|ccc|ccc|c|}
  \hline
  \multirow{2}{*}{ID} & \multirow{2}{*}{Target Function} & \multirow{2}{*}{Dist.} & \multicolumn{3}{c|}{\toolname Hit \%} & \multicolumn{3}{c|}{Syzkaller Hit \%} & \multirow{2}{*}{Avg. Diff}\\ \cline{4-9}
    &    &                   & Run 1 & Run 2 & Run 3 & Run 1 & Run 2 & Run 3 &\\ \hline
  1 & \texttt{ksys\_semctl                       } & 1 & 28.27 & 28.89 & 31.8 & 3.8 & 5.15      & 1.11    & \textbf{26.3}   \\
  2 & \texttt{\_\_sys\_setfsgid                  } & 1 & 19.1 & 13.89 & 9.45 & 0.0 & 0.03       & 0.0     & \textbf{14.14}    \\
  3 & \texttt{do\_sched\_yield                   } & 1 & 25.15 & 26.32 & 41.7 & 20.57 & 30.14   & 26.86   & \textbf{5.2}   \\
  4 & \texttt{vm\_acct\_memory                   } & 2 & 32.4 & 28.82 & 32.84 & 22.12 & 17.83   & 15.27   & \textbf{12.95}   \\
  5 & \texttt{\_\_shmem\_file\_setup             } & 2 & 8.82 & 7.32 & 7.81 & 3.79 & 6.49       & 4.93    & \textbf{2.91}   \\
  6 &\texttt{io\_register\_iowq\_m...  } & 2 & 22.05 & 15.63 & 18.26 & 1.02 & 3.28     & 2.59    & \textbf{16.35}   \\
  7 &\texttt{\_\_anon\_inode\_getfile          } & 2 & 30.47 & 30.38 & 30.65 & 9.97 & 11.89    & 11.68   & \textbf{19.32}   \\
  8 &\texttt{copy\_fsxattr\_from...         } & 3 & 56.8 & 56.99 & 54.75 & 51.0 & 49.34     & 50.1    & \textbf{6.03}   \\
  9 &\texttt{\_\_io\_uring\_add\_...    } & 3 & 36.02 & 34.97 & 28.0 & 8.9 & 2.65       & 11.76   & \textbf{25.23}   \\
  10 &\texttt{keyring\_ptr\_to\_key             } & 3 & 30.26 & 21.64 & 23.95 & 6.58 & 2.48     & 6.47    & \textbf{20.11}   \\
  11 &\texttt{mnt\_get\_writers                 } & 3 & 77.1 & 73.33 & 75.44 & 67.6 & 73.84     & 80.1    & 1.44   \\
  12 &\texttt{futex\_requeue\_pi\_...       } & 3 & 0.92 & 0.0 & 0.0 & 2.94 & 0.31          & 2.0     & -1.44   \\
  13 &\texttt{wait\_for\_device\_probe          } & 4 & 0.33 & 0.31 & 0.12 & 0.35 & 0.14        & 0.13    & \textbf{0.05}   \\
  14 &\texttt{memcpy\_to\_page                  } & 4 & 24.07 & 29.73 & 34.0 & 7.1 & 9.02       & 0.0     & \textbf{23.89}   \\
  15 &\texttt{kimage\_is\_dest...               } & 5 & 1.74 & 7.69 & 8.05 & 0.0 & 0.06         & 0.66    & \textbf{5.59}   \\
  16 &\texttt{find\_lock\_entries               } & 5 & 40.68 & 37.72 & 33.88 & 38.28 & 34.67   & 39.23   & 0.03   \\
  17 &\texttt{fsnotify\_data\_sb                } & 5 & 58.2 & 57.33 & 61.22 & 55.73 & 55.72    & 60.94   & 1.45   \\
  18 &\texttt{security\_inode\_set...         } & 5 & 12.02 & 10.74 & 12.86 & 3.39 & 4.78     & 3.61   & \textbf{7.94}   \\
  19 &\texttt{free\_partitions                  } & 6 & 13.48 & 21.94 & 14.57 & 28.17 & 24.73   & 25.97   & -9.63   \\
  20 &\texttt{bpf\_prog\_free                   } & 6 & 0.56 & 5.12 & 3.68 & 1.25 & 1.5         & 3.37    & \textbf{1.08}   \\
  21 &\texttt{locks\_delete\_glob... } & 6 & 0.59 & 0.58 & 0.0 & 0.73 & 0.04         & 0.56    & -0.05   \\
  22 &\texttt{pmd\_none\_or\_clear\_bad         } & 7 & 12.92 & 11.3 & 16.47 & 14.72 & 19.68    & 18.11   & -3.94   \\
  23 &\texttt{\_\_submit\_bio\_noac...       } & 7 & 31.89 & 21.5 & 19.88 & 20.09 & 28.69    & 27.06   & -0.86   \\
  24 &\texttt{srcu\_read\_lock\_nm...         } & 7 & 19.89 & 45.15 & 26.51 & 23.62 & 25.22   & 20.31   & \textbf{7.47}   \\
  25 &\texttt{trace\_wbc\_writepage             } & 8 & 1.82 & 0.81 & 3.03 & 0.79 & 1.71        & 0.6     & \textbf{0.85}    \\
  26 &\texttt{sk\_set\_bit                      } & 8 & 8.21 & 10.61 & 6.48 & 3.09 & 3.23       & 6.61    & \textbf{4.12}    \\
  27 &\texttt{sidtab\_search\_core              } & 8 & 76.48 & 77.85 & 75.68 & 73.14 & 76.26   & 73.34   & 2.42    \\
  \hline
  \end{tabular}
\vspace{3pt}
\caption{Experimental Data Comparison between Two Methods
(``Dist.'' denotes the minimum length of call path from some system call to target function. ``Hit \%'' represents the ratio of the test cases that covered close area in the sampled test cases in percentage. ``Avg. Diff'' denotes the average difference of the hit rate of \toolname  minus the hit rate of Syzkaller across all runs.)\label{tab:exp_table}}
\vspace{-20pt}
\end{table}

These results confirm that the LLM integration in \toolname effectively improves Syzkaller's performance in direct fuzzing, as the majority of cases achieved a higher hit rate when using \toolname.

\begin{wrapfigure}{r}{0.5\linewidth} 
    \vspace{-20pt}
    \centering
    \caption{Coverage-Execution graph for target function \texttt{sk\_set\_bit} within 2h({\color{red} red} line for Syzkaller and {\color{blue} blue} line for \toolname)}
    \includegraphics[width=\linewidth]{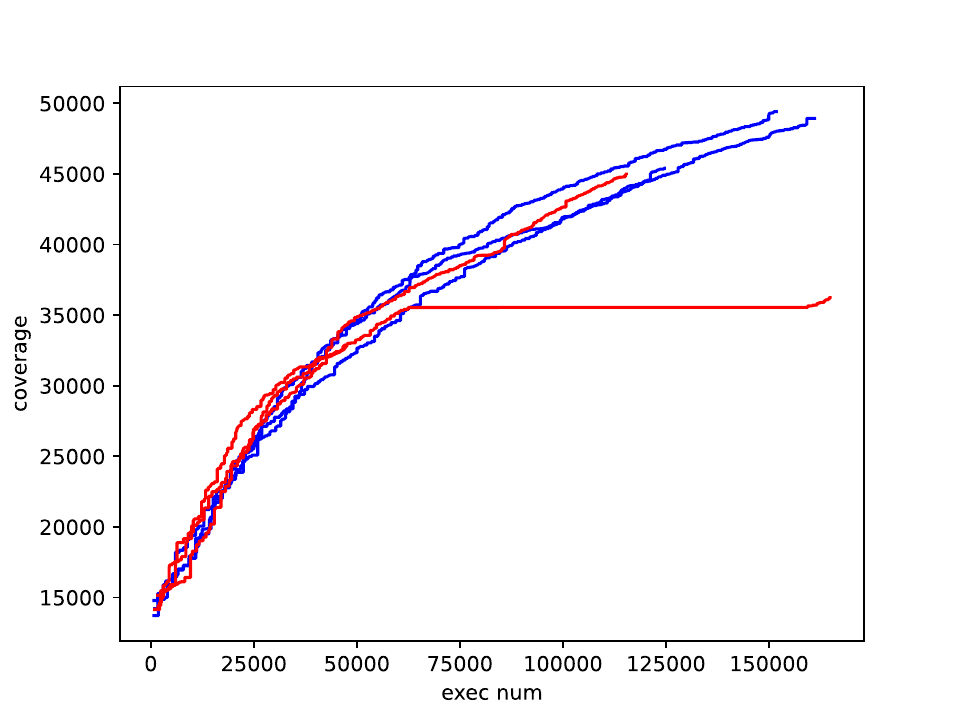}
    \label{fig:break}
    \vspace{-30pt}
\end{wrapfigure}

While this paper primarily focuses on kernel direct fuzzing, during our experiments, we observed that \toolname successfully breaks the Syzkaller coverage plateau. In the 27 direct fuzzing cases, we found that 5 cases achieved higher coverage within a fixed number of test cases, with IDs 4, 19, 25, 26, 27. Figure~\ref{fig:break} illustrates the coverage progression for case 27, where deeper target functions were tested, partially validating our hypothesis that the main reason of plateau is the fuzzer lacking a seed that can reach deeper code. However, we also noted that in 6 cases, the coverage performance of \toolname was inferior to that of Syzkaller, while the remaining cases showed similar performance between the two tools. We identify this as another promising new direction emerging from this work, and it will be valuable to investigate this hypothesis further, exploring how to harness the LLM's capabilities to systematically improve kernel fuzzing coverage.

\begin{table}[]
    \centering
    \begin{tabular}{c|c|c|c}
        ID & Target Function & \toolname  & Syzdirect \\
        \hline
        1 & \texttt{ stable\_page\_flags} &\makecell{\texttt{read, pread}} & \makecell{\texttt{{\color{red}ioctl(4 variants}),} \\ {\texttt{{\color{teal}io\_uring\_enter}, read,}} \\ \texttt{{\color{red}write}, mount}} \\
        \hline
        2  & \texttt{fscontext\_create\_fd}&\makecell{\texttt{ \color{teal}fsopen, fspick}} & \makecell{ \color{teal}None} \\
        \hline
        3 &\texttt{memfd\_fcntl}
          & \makecell{\texttt{fcntl, { \color{teal}fcntl64},}\\ \texttt{ioctl(UDMABUF\_CREATE)}} & \makecell{\texttt{ioctl(UDMABUF\_CREATE,}\\\texttt{{ \color{red} UDMABUF\_CREATE\_LIST}), }\\\texttt{fcntl({\color{red}getflags})}} \\
    \end{tabular}
    \vspace{5pt}
    \caption{Exemplar system call entry analysis results from \toolname and SyzDirect reveal that \toolname has advantages over SyzDirect in identifying system call relationships, as highlighted in {\color{teal}green}. Conversely, SyzDirect excels in detecting argument types, as shown in {\color{red}red}, a feature not currently supported by \toolname.}
    \label{tab:static_analysis}
    \vspace{-20pt}
\end{table}

\paragraph{\toolname vs SyzDirect}
An end-to-end comparison with the SyzDirect tool was not feasible due to multiple issues encountered during its installation, configuration, and manual instrumentation requirements\footnote{Despite our efforts to resolve these issues, we were unable to execute its fuzzing process due to unresolved configuration errors, and although we reached out to its authors, we did not receive a response.}. 

Nevertheless, we managed to run SyzDirect's stages for system call entry analysis and conducted a comparison with the LLM-generated results from \toolname. Table~\ref{tab:static_analysis} presents the results for three target functions in the Linux kernel \footnote{Commit 304040fb4909f7771caf6f8e8c61dbe51c93505a}. In the table, system calls highlighted in {\color{teal} green} indicate cases where \toolname outperforms SyzDirect, while those in {\color{red} red} represent cases where SyzDirect performs better.

In cases with IDs 2 and 3, \toolname identified three additional system call entries compared to SyzDirect. After manually verifying these cases, we found that SyzDirect's call graph analysis was less precise than that of \toolname. For example, in the first case, 
\texttt{io\_uring\_enter} did not appear to be beneficial for reaching the target function. However, SyzDirect outperformed \toolname in providing specific variants of system calls, likely due to its more detailed call graph model that incorporates resource-producing and consuming relationships, which are currently not included in \toolname analysis.  
This results in a finer-grained analysis by SyzDirect compared to that of \toolname.

\section{Conclusion and Discussion}
In this work, we explored the integration of LLM capabilities with OS kernel fuzzers in real-time. Based on our preliminary experimental results, this approach appears effective for direct fuzzing and warrants further investigation. However, our work is still in its early stage, as several advanced techniques, such as the relational graph approach from \cite{healer} and more sophisticated static analysis methods like those in \cite{statefuzz, difuze}, have not yet been incorporated. Our work also lacks the validation on whether the system calls are correctly generated.

At the implementation level, there are several ways \toolname could be enhanced:
\begin{inparaenum}[1)]
    \item Splitting the calling code into smaller segments to facilitate deeper exploration of target functions ;
    \item Integrating more closely with Syzkaller to enable LLMs to contribute to argument mutation processes; and
    \item Using the distance to the target function of cases that cover nearby areas to select the most promising test cases for generating feedback prompts.
\end{inparaenum}

We regard LLMs as a viable solution to the complexities inherent in OS kernel fuzzing, thanks to the vast amount of data on which they are trained and optimized. The combination of LLM capabilities with our real-time feedback framework offers a flexible way to automatically adjust the fuzzing strategy. In the future, we believe it will be important to continue researching how LLMs can boost fuzzing coverage by utilizing information from intermediate results of static analysis and kernel documentation.

\begin{credits}
\subsubsection{\ackname} We gratefully thank  Pierre Olivier  for providing insights of linux kernel on this study. This work is partly supported by CAS Project for Young Scientists in Basic Research, Grant No.YSBR-040, ISCAS New Cultivation Project ISCAS-PYFX-202201, ISCAS Basic Research ISCAS-JCZD-202302 and the Ministry of Education, Singapore under its Academic Research Fund Tier 3 (Award ID: MOET32020-0004).

\end{credits}
%
%
%
\bibliographystyle{splncs04}
\bibliography{references}

\end{document}